\documentclass{article}
\usepackage{graphicx}
\usepackage{epsfig}
\usepackage{amsmath}

\newcommand{\spone}{0.9}  

\newcommand{\sptwo}{1.4}
\newcommand{\spthree}{2.4}

\newcommand{\singlespace}{\edef\baselinestretch{\spone}\Large\normalsize}

\newcommand{\doublespace}{\edef\baselinestretch{\sptwo}\Large\normalsize}
\newcommand{\threespace}{\edef\baselinestretch{\spthree}\Large\normalsize}

\singlespace
\threespace
\doublespace

\everymath={\displaystyle}
\begin{document}
\doublespace

\begin{center}
{\bf {\large Nonlinear Realization and Weyl Scale Invariant p=2 
Brane}\\
$~$\\
Lu-Xin Liu\\
{\it Department of Physics, Purdue University,\\ 
West Lafayette, IN 47907, USA\\
liul@physics.purdue.edu \\}}
\end{center}
\vspace{25pt}
{\bf Abstract.} The action of Weyl scale invariant p=2 brane which breaks 
the target super Weyl scale symmetry in the N=1, D=4 superspace down to 
the lower dimensional Weyl symmetry $W(1,2)$ is derived by the approach of 
nonlinear realization. The dual form action for the Weyl scale 
invariant supersymmetric D2 brane is also constructed. The interactions of 
localized matter fields on the brane with the Nambu-Goldstone fields associated 
with the breaking of the symmetries in the 
superspace and one spatial translation directions are obtained through the 
Cartan one-forms of the Coset structures. The covariant 
derivatives for the localized matter fields are also obtained by 
introducing Weyl gauge field as the compensating field 
corresponding to the local scale transformation on the brane world volume.  
\vspace{60pt}

\pagebreak

\begin{flushleft}
{\large I. Introduction}
\end{flushleft} 
\vspace{5pt}      

   Nonlinear realization of a compact Lie group can be realized on the 
Nambu-Goldstone fields related to the broken 
symmetry generators and it becomes linear when restricted to the given 
subgroup [1]. In Ref.[2], based on nonlinear 
transformation of a spinor field, which played a role of Goldstino field, 
the nonlinear realization method was extended 
to include fermion like generators. The resulted degeneracy of vacuum gave 
rise to spontaneous broken of supersymmetry. 
The approach of nonlinear realization of SUSY, besides the Goldstone field, 
was generalized to matter fields as well as 
gauge fields [3,4], with a formalism of effective couplings to the 
Goldstino field. In Ref.[5], one can find 
applications of nonlinear realization to branes of M theory, and a general 
description is given to derive the dynamics 
of the branes. There, it is restricted to group $G$ whose generators can 
be divided into two subgroups with one (such as 
Lorentz group) is the automorphism group of another (whose generators 
associated with (super)spacetime positions).  The 
transformation of the group $G$ with respect to the coset of the unbroken 
automorphism generators group would give us a 
description of the embedded submanifold, which has the dimensions of the 
coset space of the unbroken automorphism 
generators group with respect to the unbroken subgroup.

    As presented in [6-10], the approach of nonlinear realization was 
extensively used to describe the spontaneous 
partial breaking of (extended) supersymmetry and construct actions of 
(super)brane dynamics. On the other hand, when 
considering conformal transformation, in Ref.[11] the dynamics of conformally 
invariant p-branes was introduced. In 
[12,13], it was further extended to describe Weyl invariant D $p$ -brane 
and superconformal supermembrane. It is our 
purpose of this paper to introduce a Weyl scale (due to dilatation 
operator $D$) 
invariant $p$=2 brane which embedded in the 
target N=1, D=4 superspace defined by \{$x_\mu  ,\theta ,\bar \theta$\}. 
Considering the unbroken subgroup $W(1,2)$  
(Weyl group) of the super Weyl group $G$ and the coset with respect to the 
unbroken automorphism group of the unbroken 
subgroup, the symmetry $G$ can be realized on the nonlinear transformation 
of collective coordinates fields which is a 
result of acting a group element of $G$ on the coset representative 
element $\Omega$. 
When applied to brane theory, it is 
illustrated that for such a brane that breaks the supersymmetry and one 
spatial translation symmetry, its dynamics is 
described by the low energy oscillations of the Nambu-Goldstone modes 
associated with these broken symmetries. 
Accordingly, the invariant action of the brane can be obtained by using  
vielbein and connection one forms on the 
submanifold after constructing Cartan one-forms from $ \Omega ^{ - 1} 
d\Omega $.

    In this paper, we start from introducing the super Weyl scale group 
and its automorphism subgroup, then use the above 
stated formalism of the nonlinear realization approach to find the 
fluctuations modes of Goldstone bosons (Goldstino 
fermions) associated with spacetime coordinates (Grassmann coordinates) of 
the broken symmetry (supersymmetry). A Weyl 
scale invariant p=2 brane will be given when the target D=4 superspace 
\{$x_\mu  ,\theta ,\bar \theta$\} broken down to D=3 spacetime world 
volume described by parameters \{$x_0  ,x_1,x_2$\} in the static gauge 
with Weyl 
scale (dilatation) symmetry 
kept, which becomes a local symmetry on the 
p=2 brane world volume. The dual form non-BPS Weyl scale invariant D2
brane supersymmetric Born-Infeld action is also obtained.

    Finally, in addition to the massless Nambu-Goldstone fields of the p=2 
brane oscillations, we also consider matter 
fields degree of freedom localized on the domain wall brane. The Weyl scale 
invariant actions of these matter fields 
are constructed by using Weyl gauge field and spin connections. The latter 
gives interactions of the matter fields with 
the Nambu-Goldstone fields.
  
\vspace{5pt}
\begin{flushleft}
{\large II. Weyl Scale Invariant p=2 Brane}
\end{flushleft}
\vspace{5pt}

    Consider super-Weyl group $G$, whose generators include N=1, D=4 
super-Poincare generators ($P_\mu,M_{\mu\nu}$,two Weyl spinor 
supersymmetry charges $Q_\alpha$,$\bar Q_{\dot \alpha}$) and the Weyl 
scale (dilatation) generator $D$.  It 
has the following (anti)commutation 
relations 
\begin{align*}
\{ Q_\alpha  ,\bar Q_{\dot \alpha } \} &= 2\sigma _{\alpha
\dot \alpha }^\mu  P_\mu \\
[M_{\mu \nu } ,M_{\rho \sigma } ] &= i(\eta _{\mu \sigma } M_{\nu \rho }
+ \eta _{\nu \rho } M_{\mu \sigma }  - \eta
_{\mu \rho } M_{\nu \sigma }  - \eta _{\nu \sigma } M_{\mu \rho } ) \\
[Q_\alpha  ,M_{\mu \nu } ] &= i\frac{1}{2}(\sigma _{\mu \nu } )_
\alpha  ^{\hspace{3pt}\beta}  Q_\beta \\
[\bar Q^{\dot \alpha } ,M_{\mu \nu } ] &= i\frac{1}{2}(\bar \sigma _
{\mu \nu } )^{\dot \alpha } _{\hspace{3pt}\dot \beta } \bar
Q^{\dot \beta }  
\tag{1}
\end{align*}
and
$$ [D,M_{\mu \nu } ] = 0,
 [D,D] = 0,
 [D,P_\mu  ] =  - iP_\mu   \\
  \eqno{(2)}
$$
$$ [D,Q_\alpha  ] =  - \frac{1}{2}iQ_\alpha,
 [D,\bar Q_{\dot \alpha } ] =  - \frac{1}{2}i\bar Q_{\dot \alpha },  \\
  \eqno{(3)}
$$
where the dilatation operator $D = -ix^\mu  \frac{\partial }{{\partial 
x^\mu}}$ in $x$-representation. From Eqs.(1-3), one can
find generators $\{M_{\mu\nu}, D\}$ form a subgroup $H^\prime$ which is 
the 
automorphism group of
another subgroup, i.e. super spacetime group by the
set of charges  of $\{ Q_\alpha  ,\bar Q_{\dot \alpha } ,P_\mu  \}$ . In 
the case when the $G$ group symmetry is broken 
to the $1+2$ dimensional Weyl $W(1,2)$ symmetry [14], whose
unbroken generators, for example in the static gauge, are $\{ M_{ij} 
,D,P_i \} $, where the
index $i=0,1,2$, and the spontaneously broken
automorphism generators are $M_{i3}$.  In such a case, we have a two 
dimensional brane which is embedded in the superspace and
breaks down the target space super Weyl invariance to a lower dimensional
Weyl group symmetry $W(1,2)$. Besides $M_{i3}$, the broken
generators in superspace are the generators $Q_\alpha  ,\bar Q_{\dot 
\alpha }$ in the Grassmann coordinate
directions  $\{x_\mu  ,\theta _\alpha  ,\bar \theta _{\dot \alpha }\}$   
and the translation generator $P_3$
transverse to the brane.

     Consider the coset $G/H$, where $H$ has unbroken antomorphism 
generators $\{M_{ij},D\}$. We hence have a p=2 brane, which has a
$W(1,2)$ symmetry, moving through the coset space $G/H^\prime$ with 
tangent group $H^\prime$. It sweeps out
a submanifold that has the dimensions of
the coset space $G^\prime/H$ with tangent group $H$, where 
$G^\prime$ is 
spanned by the unbroken
automorphism generators $\{M_{ij},D\}$ and the unbroken
spacetime generators $P_i$. Since we will work on a D=3 manifold, it is 
more convenient to express N=1,D=4 super Weyl algebra in terms of D=3 
Lorenz group indices with new defined 
generators $M^m  = \frac{1}{2}\varepsilon ^{mnr} M_{nr} $ and 
$K^m=M^{m3}$, where $m=0,1,2$.  In N=2, D=3 supersymmetry
theory, there are four supercharges which are the same number as for N=1
supersymmetry in D=4. In fact, N=2, D=3 SUSY
algebra [15-17] can be obtained by dimensionally reducing N=1
supersymmetry in four dimensions. From the D=3
standpoint the N=1, D=4 SUSY algebra is a central-charged extended N=2
Poincare superalgebra, with one D=4 translation
generator becomes the central charge generator $Z_0$ [8,10,15]. Taking    
$Z_0=-P_3=-Z$, with supercharges $Q_\alpha  ,\bar Q_{\dot \beta }$ 
becomes two complex
conjugate spinors $Q_\alpha  ,\bar Q_{\beta }$ in three dimensions, 
the N=1, D=4 Supersymmetry 
reduces to N=2, D=3 extended Supersymmetry with
relations (See Appendix A for derivation and notations):
$$ [M^m ,M^n ] =  - i\varepsilon ^{mnr} M_r ,
 [M^m ,K^n ] =  - i\varepsilon ^{mnr} K_r,
 [K^m ,K^n ] =  i\varepsilon ^{mnr} M_r  \\$$
$$[M^m ,P^n ] =  - i\varepsilon ^{mnr} P_r,
 [M^m ,Z] = 0,
 [K^m ,P^n ] =  + i\eta ^{mn} Z,
 [K^m ,Z] =  + iP^m  \\$$
$$[Z,P_\mu  ] = 0,
 [D,q_i ] =  - \frac{1}{2}iq_i,
 [D,s_i ] =  - \frac{1}{2}is_i  \\$$
$$\{ q_i ,q_j \}  = 2(\gamma ^m C)_{ij} P_m,
 \{ s_i ,s_j \}  = 2(\gamma ^m C)_{ij} P_m,
 \{ q_i ,s_j \}  =  - 2iC_{ij} P_3  \\
  \eqno{(4)}
$$
where $q,s$ are extended supercharges in three dimensions. The unbroken
automorphism generators forms group $H$. With R
symmetry suppressed, an exponential description of the Coset $G/H$  
representative
element is
$$  \Omega  = e^{i\xi ^m p_m } e^{i[\phi (\xi )Z + \bar \theta _i (\xi 
)q_i
+ \bar \lambda _i (\xi )s_i ]} e^{iu^m
(\xi )K_m },  \\
  \eqno{(5)}
$$
in which variables $\xi$ parameterizes the embedded submanifold described 
by the p=2 brane, and $\phi (\xi ),\theta (\xi ),\lambda (\xi ),u(\xi)$ 
are the Nambu-Goldstone fields that depend on variables $\xi$. The 
dynamics can be
constructed about the brane which describe a
broken symmetry in the $z,\theta ,\bar \theta $ superspace coordinates 
directions and whose
long wave length excitation modes are described by
these Nambu-Goldstone fields associated with these broken symmetries.
By using reparameterization invariance, we choose
static gauge $x^m=\xi^m$  for space time coordinates $x^m$ lying in 
directions of the
brane. Then it becomes
$$\Omega  = e^{ix^m p_m } e^{i[\phi (x)Z + \bar \theta_i (x) q_i  + \bar
\lambda_i (x)s_i ]} e^{iu^m (x)K_m }.  \\
  \eqno{(6)}
$$
The elements of group $G$ can be decomposed uniquely into a product form 
of
Coset representative element $\Omega$ and subgroup
element of $H$. In some neighborhood of the identity of group G, its 
element is parameterized as
$$  g = e^{i[a^m p_m  + \bar \xi q + \bar \eta s + zZ + b^m K_m  + \alpha 
^m M_m  + dD]}  \\
  \eqno{(7)}
$$
Under a right group transformation $g$, the Coset element $\Omega$   
transforms to\\
$\Omega^\prime = e^{ix^{\prime m} p_m } e^{i[\phi^{\prime} 
(x^\prime)Z + \bar \theta 
^\prime_i (x^\prime) q_i  + \bar \lambda ^\prime_i (x^\prime)s_i ]} 
e^{iu^{\prime m} (x^\prime)K_m } $
with the following relation
$$  g\Omega  = \Omega^\prime h, \\
  \eqno{(8)}
$$
where $h$ stands for the subgroup element. The field $\phi(x)$, which 
transforms
linearly under rigid $g$ transformations, i.e. $ \phi^\prime (x) = g\phi 
(x)$, could be re-expressed as field $\tilde \phi (x) $ through 
$\Omega ^{ - 1} \phi (x) = \tilde \phi (x)$ from 
which the massless 
Goldstone
mode has been eliminated. Therefore, when
the symmetry group $G$ is broken to subgroup $H$, from Eq.(8) one finds 
that $\phi^\prime (x) = \Omega^\prime \tilde \phi^\prime (x) = g\phi (x) 
= g\Omega \tilde 
\phi (x) = \Omega^\prime h\tilde \phi (x)$.
Then the original $G$ transformations is
re-written as the transformation depending on $\tilde 
\phi^\prime (x) =h \tilde \phi (x)$ under the unbroken subgroup $H$,
which is used as basic formalism to
construct invariant Lagrangian when considering localized matter fields on
the brane [18]. Applying Eqs.(B1)-(B3),  it can
be found the transformations of the space coordinates as well as the
Nambu-Goldstone fields induced by the
infinitesimal transformation of group $g$:
 $$  x'^m  = x^m  + dx^m  + a^m  - i(\bar \xi \gamma ^m \theta  + \bar 
\eta
\gamma ^m \lambda ) - \phi b^m  +
\varepsilon ^{mnr} \alpha _n x_r,  \\$$
$$ \phi '(x') - \phi (x) = \Delta \phi  = z + d\phi  + (\xi \gamma ^0
\lambda  - \theta \gamma ^0 \eta ) - b^m x_m,  \\$$
$$ \theta '(x') - \theta (x) = \Delta \theta _i  = \xi _i  + \frac{1}{2}d
\theta _i  +\frac{i}{2}b_m (\gamma ^m \lambda
)_i  - \frac{i}{2}\alpha _m (\gamma ^m \theta )_i,  \\$$
$$ \lambda '(x') - \lambda (x) = \Delta \lambda _i  = \eta _i  + 
\frac{1}{2}
d\lambda _i  -\frac{i}{2}b_m (\gamma ^m
\theta )_i  - \frac{i}{2}\alpha _m (\gamma ^m \lambda )_i,  \\$$
$$ u'(x') - u(x) = \Delta u^m  = \frac{{\sqrt {u^2 } }}{{\tanh \sqrt {u^2 
} }}(b^m  - \frac{{u^r b_r u^m }}{{u^2 }}) +
\frac{{u^r b_r u^m }}{{u^2 }} + \varepsilon ^{mnr} \alpha _n u_r,  \\
  \eqno{(9)}
$$
where the linear terms of $d$ represent the Weyl scale transformations of
each collective coordinates. The element $h$ is
given by
  $$ h = e^{i[\alpha ^m M_m  -\frac{1}{2}\frac{{\tanh \frac{1}{2}
\sqrt {u^2 } }}{{\frac{1}{2}\sqrt {u^2 } }}b_n u_r
\varepsilon ^{nrm} M_m  + dD]}  \\
  \eqno{(10)}
$$
From above, it can be found the spacetime coordinates have a field 
dependent
transformation as a result of the
nonlinear realization of group $G$. In this case, there are broken 
symmetries of $Q_\alpha  ,\bar Q_{\dot \alpha } ,Z $ and 
rotation generators $M^{m3}$ related to the $z$
direction. For the breaking symmetry of spacetime, the only 
Nambu-Goldstone
fields are those associated to the broken
(super)translations [19], and the superfluous Nambu-Goldstone fields
$u^m$ can be eliminated by imposing invariant
conditions on  the Cartan differential forms [20] (see equation (19)).

     The $G$ symmetry is represented by transformation properties of the 
field $\tilde \phi (x) $ under the unbroken subgroup $H$. Considering
incorporation of the dynamics of the field $\tilde \phi (x) $  with that 
of the brane, we
work on dreibein basis in the local tangent
space of the submanifold swept out by the p=2 brane. The interval
$ ds^2  = g_{mn} dx^m dx^n $ has
the form $ ds^2  = \eta_{ab} dx^a dx^b $ in the tangent space, with
relations $ ds^2  = g_{mn} dx^m dx^n = \eta_{ab}e_m^{\\\ a} e_n^{\\\ b} 
dx^m dx^n $ and 
$dx^a=e_m^{\\\ a} dx^m$. 
The metric tensor is related to the dreibein through
$$ g_{mn}  = e_m^{\\\ a} e_n^{\\\ b} \eta _{ab}  \\
  \eqno{(11)}
$$       
In the local subgroup $H$ formed by algebra $\{M_{ij},D\}$, under the 
scale transformation  $x^m  \to x^{\prime m}  = e^d x^m $,
the interval transforms as $ds^2  \to ds^{\prime 2}  = e^{2d} ds^2 $. In 
terms of metric tensor $g_{mn}$, because $ds^2  = g_{mn} dx^m dx^n$, 
then it is understanding that the metric tensor have
a weight 2 under the Weyl scale transformation,
i.e. 
$$ g_{mn}^\prime = e^{2d} g_{mn}.  \\
  \eqno{(12)}
$$    
Conversely, $g^{mn}$ has scale weight -2. Its total infinitesimal 
transformation induced by the general coordinate
variation $x^m  \to x^{\prime m}  = x^m  + dx^m +\varepsilon ^m (x) $ is 
given by
$$ g_{mn}^\prime = g_{mn}+2dg_{mn}  - (\partial _m \varepsilon _n  + 
\partial _n
\varepsilon _m ) \\
  \eqno{(13)}
$$
where $\varepsilon ^m (x) = a^m  - i(\bar \xi \gamma ^m \theta  
+ \bar \eta \gamma ^m \lambda ) - \phi b^m  + \varepsilon ^{mnr} \alpha 
_n x_r  $.

     In order to construct an invariant action, we can obtain dreibein
and connection one-forms by using Cartan form
$\Omega^{-1}d\Omega$, which is expanded with respect to the
$G$ generators:
$$ \Omega ^{ - 1} d\Omega  = i(\omega ^a p_a  + \bar \omega _{qi} q_i  +
\bar \omega _{si} s_i  + \omega _Z Z + \omega _k^a K_a  + \omega _M^a M_a
+ \omega _D D) \\
  \eqno{(14)}
$$
Under the transformation $\Omega \to \Omega^\prime$, the Cartan forms 
transform as
$$  \Omega ^{\prime - 1} d\Omega^\prime = h(\Omega ^{ - 1} d\Omega )h^{ - 
1}  +
hdh^{ - 1}.  \\
  \eqno{(15)}
$$
It is obvious that all the forms transform homogeneously under $G$ except 
the connection one from $\omega^a_M$ which transforms by a
shift. These forms are invariant under the global left action of $G$ on 
$G/H$ . Under the local right action $\Omega  \to \Omega^\prime h $  with 
$h$ given by Eq.(10), the forms $\omega^a$ transform as the dreibein on 
the tangent bundle to $G/H$,
while $\omega^a_M$ transforms as a connection to this
bundle. The Cartan forms associated with the unbroken spacetime generators
$P$ involve the exterior derivative $d$ which is
independent of the coordinate system used to parameterize the embedded
submanifold and is reparameterization
invariant.  After choosing the static gauge $\xi^m=x^m$, the dreibein 
$e^{\\\ a}_m$ is obtained by
expanding spacetime one-forms $\omega^a$ with respect
to the coordinate differentials $dx^m$, i.e. $\omega^a=dx^m e^{\\\ a}_m$.  
The connection one-forms $\omega^a_M$, on
the other hand, can be used to construct the
covariant derivative of the fields
$ \nabla \tilde \phi (x) = (d + i\omega _M^a \Gamma _a  + i\omega _D 
\Gamma^\prime)\tilde \phi (x) $, where $\Gamma_a$ and $\Gamma^\prime$ are 
respective representations of the generators $M_a$ and
$D$ with respect to the fields $\tilde \phi(x)$. These are the
building blocks that can be used to construct invariant actions under $G$.
Considering equations (4) and (B4), we have
\begin{align*}
  \omega _M^a  &= (\cosh \sqrt {u^2 }  - 1)\frac{{u_b du_c }}{{u^2
}}\varepsilon ^{abc}  \\
 \omega ^a  &= (dx^m  + id\theta \gamma ^0 \gamma ^m \theta  + id\lambda
\gamma ^0 \gamma ^m \lambda ) \\
  &\cdot (\delta _m ^{\hspace{3pt} a}  + (\cosh \sqrt {u^2 }  - 
1)\frac{{u_m u^a
}}{{u^2 }}) + (d\phi  + d\theta \gamma ^0 \lambda  - d\lambda 
\gamma ^0 \theta )\frac{{\sinh \sqrt {u^2 } }}{{\sqrt {u^2 } }}u^a  \\
\omega _D  &= 0 
\tag{16}
\end{align*}
where $a=0,1,2$. We use $a,b,c$ to represent the tangent spacetime index, 
and $i,j,k$ to 
represent 2+1 general coordinates in what follows.
Since $D$ is the automorphism generator of the (super)spacetime position 
group, and from the commutation relations of
Eq.(2), it is found that $D$ is not involved in the Cartan forms here, 
which is in consistent with $\omega_D=0$. The dreibein
\begin{align*}
e_m^{\hspace{3pt} a} & = (\delta _m^{\hspace{3pt} b}  + i\partial _m 
\theta \gamma ^0 
\gamma ^b
\theta  + i\partial _m \lambda \gamma ^0 \gamma ^b \lambda ) \\
  &\cdot (\delta _b^{\hspace{3pt} a}  + (\cosh \sqrt {u^2 }  - 
1)\frac{{u_b u^a
}}{{u^2 }} + (\tilde D_b \phi  + \tilde D_b \theta \gamma ^0 \lambda  -
\theta \gamma ^0 \tilde D_b \lambda )\frac{{\sinh \sqrt {u^2 } }}{{\sqrt
{u^2 } }}u^a ) \\
&={A^{\hspace{3pt} b}_m} \cdot (\delta _b^{\hspace{3pt} a}  + (\cosh 
\sqrt {u^2 }  - 1)\frac{{u_b 
u^a}}{{u^2 }} + (\tilde D_b \phi  + \tilde D_b \theta \gamma ^0 \lambda  -
\theta \gamma ^0 \tilde D_b \lambda )\frac{{\sinh \sqrt {u^2 } }}{{\sqrt
{u^2 } }}u^a ) 
  \tag{17}
\end{align*}
has a tangent space index $a$, which has the transformation property 
induced by Eq.(15) in the local tangent space, with $L^{\\\ a}_b$
the local $H$ representation with vector indices
$$  e_m^{\prime \\\ a}  = e_m^{\\\ b} L_b^{\\\ a}  \\
  \eqno{(18)}
$$
In Eq.(17), $\tilde D_b =A_b^{-1m}\partial_m$ is the Akulov-Volkov 
derivative, defined by $A_m ^{\\\ b}  = \delta _m^{\\\ b}  + i\partial _m 
\theta \gamma ^0 \gamma ^b \theta  + i\partial _m \lambda \gamma ^0 \gamma 
^b \lambda  $[3,4,10]. Imposing the invariant condition $\omega_z=0$ on 
the covariant derivative, as a result of the inverse Higgs Mechanism 
[10,20], the field $u_m$ can be eliminated by the following relation
  $$  u_b \frac{{\tanh \sqrt {u^2 } }}{{\sqrt {u^2 } }} =  - (\tilde D_b
\phi  + \tilde D_b \theta \gamma ^0 \lambda  - \theta \gamma ^0 \tilde D_b
\lambda ) =  - \tilde D_b \Phi  \\
  \eqno{(19)}
$$
Plugging this into Eq.(17), the dreibein hence has the simple form
\begin{align*}
e_m ^{\hspace{3pt} a}  &= A_m ^{\hspace{3pt} b}  \cdot (\delta 
_b^{\hspace{3pt} a}  + \frac{{u_b u^a }}{{u^2
}}(\frac{1}{{\cosh \sqrt {u^2 } }} - 1)) \\
&= A_m ^{\hspace{3pt} b}  \cdot (\delta _b^{\hspace{3pt} a}  + 
\frac{{\tilde D_b 
\Phi \tilde D^a \Phi }}{{(D\Phi )^2 }}(\sqrt {1 - (\tilde D\Phi )^2 }  - 
1)). 
  \tag{20}
\end{align*}
The metric tensor becomes
  $$ g_{mn}  = e_m^{\\\ a} e_n^{\\\ b} \eta _{ab}  = A_m ^{\\\ a} A_n 
^{\\\ b} \eta _{ab}  
- \partial _m \Phi \partial _n \Phi  \\ 
  \eqno{(21)}
$$
Introduce four dynamic variables $X^\mu=(X^a,X^3)=(X^a,\Phi)$, which are 
defined as following
\begin{align*}
dX^a  &= dx^m A_m^{\hspace{3pt} a}  \\
dX^3  &= (\partial _m \phi  + \partial _m \theta \gamma ^0 \lambda  -
\theta \gamma ^0 \partial _m \lambda )dx^m 
\tag{22}
\end{align*}
After integrating from both sides, we have
\begin{align*}
X^0  &= x^0  + f^0 (\theta ,\lambda ), \\
X^1  &= x^1  + f^1 (\theta ,\lambda ), \\
X^2  &= x^2  + f^2 (\theta ,\lambda ), \\
X^3  &= \phi  + F(\theta ,\lambda ), 
\tag{23}
\end{align*}
where $f(\theta,\lambda),F(\theta,\lambda)$ are functions of 
$\theta(x),\lambda(x)$, decided by the integration of Eq.(22). 
Therefore,in the static gauge $\xi^m=x^m$, by using Eq.(23), the metric 
tensor in
Eq.(21) now becomes
  $$  g_{mn}  = \eta _{\mu \nu } \frac{{\partial X^\mu  }}{{\partial \xi
^m }}\frac{{\partial X^\nu  }}{{\partial \xi ^n }} = \eta _{\mu \nu }
\frac{{\partial x^\mu  }}{{\partial \xi ^m }}\frac{{\partial x^\nu
}}{{\partial \xi ^n }}+other \\\ terms \\
\eqno{(24)}
$$                                
where $x^\mu=(x^0,x^1,x^2,\phi)$. Consequently, in contrast with the 
normal spacetime induced metric $g_{mn}  = \eta _{\mu \nu } 
\frac{{\partial x^\mu  }}{{\partial \xi
^m }}\frac{{\partial x^\nu  }}{{\partial \xi ^n }}$ on
p brane world volume, there are modification terms
of the metric through the functions $f(\theta,\lambda),F(\theta,\lambda)$ 
which are contributed from the Nambu-goldstone
fields $\theta(x),\lambda(x)$ corresponding to the broken
symmetries in the superspace coordinates directions (see equation (28) 
for details).

     The infinitesimal transformation of dreibein in the local tangent 
space is
$$  \delta e_m^{\\\ a}  = \delta ^L e_m^{\\\ a}  + de_m^{\\\ a}  \\
\eqno{(25)}
$$
with $\delta ^L e_m^a$ represents the local Lorentz transformation. The 
second term is the Weyl scale transformation. Hence, the world
volume has the scale transformation property
  $$ dx^3 \det e \to dx'^3 \det e' = e^{ 3d} dx\det e. \\
  \eqno{(26)}
$$
We introduce the intrinsic metric $\rho_{mn}$ on this p=2 brane manifold. 
Similarly, it has the Weyl scale transformation property $\rho_{mn} \to 
e^{2d}\rho_{mn}$ and $dx^3 \sqrt{|\rho|} \to dx'^3 \sqrt{|\rho'|} = 
e^{3d} dx^3 \sqrt{|\rho|}$. Considering Eq.(24), the action of the Weyl 
scale invariant p=2 brane is then constructed 
$$
I =  - T\int {d^3 x} \sqrt{|\rho|} (\frac{1}{3}\rho^{mn} \eta _{\mu \nu 
} \partial _m X^\mu  \partial _n X^\nu  )^{3/2}
\eqno{(27)}
$$
here $\rho^{mn}$ is the inverse of the metric $\rho_{mn}$ and $\rho$ 
stands for the determinant of $\rho_{mn}$. The auxiliary intrinsic metric   
$\rho_{mn}$ can be eliminated by using its equation 
of motion. By using Eq.(22), the action then has an explicit form
\begin{align*}
I &=  - T\int {d^3 x} \sqrt{|\rho|} (\frac{1}{3}\rho^{mn} \eta _{\mu \nu
} \partial _m X^\mu  \partial _n X^\nu  )^{3/2}\\
&= - T\int {d^3 x} \sqrt{\rho} \{ \frac{1}{3}\rho^{mn} (\eta _{mn}
+ i\partial _n \theta \gamma ^0 \gamma _m
\theta  + i\partial _n \lambda \gamma ^0 \gamma _m \lambda  + i\partial _m
\theta \gamma ^0 \gamma _n \theta  + i\partial _m \lambda \gamma ^0 \gamma
_n \lambda \\
&+ (i\partial _m \theta \gamma ^0 \gamma ^a \theta  + i\partial _m \lambda
\gamma ^0 \gamma ^a \lambda ) \cdot (i\partial _n \theta \gamma ^0 \gamma
_a \theta  + i\partial _n \lambda \gamma ^0 \gamma _a \lambda ) \\
&-(\partial_m \phi+\partial _m \theta \gamma ^0 \lambda - 
\theta \gamma^0 \partial _m \lambda) \cdot
(\partial_n \phi+\partial _n \theta \gamma ^0 \lambda -
\theta \gamma^0 \partial _n \lambda)) \}^{3/2}
\tag{28}
\end{align*}                                         
where $T$ stands for the brane tension.

\vspace{5pt}
\begin{flushleft}
{\large III. Weyl Scale Invariant D2 Brane}
\end{flushleft}
\vspace{5pt}

     As discussed in section II, we have constructed the Weyl scale 
invariant non-BPS p=2 brane action. In the following we derive its dual 
form, the non-BPS Weyl scale invariant D2 brane supersymmetric Born-Infeld 
action. From Eqs.(19) and (20), we have
\begin{align*}                       
\det e &= \det (e_m ^{\hspace{3pt} a} ) = \det \{ A_m ^{\hspace{3pt} b}  
\cdot (\delta 
_b^{\hspace{3pt} a}  + 
\frac{{u_b u^a }}{{u^2 }}(\frac{1}{{\cosh \sqrt {u^2 } }} - 1))\} \\ 
&= \det (A_m ^{\hspace{3pt} b} ) \cdot \det \{ \delta _b^{\hspace{3pt} a}  
+ \frac{{u_b 
u^a }}{{u^2 }}(\frac{1}{{\cosh \sqrt {u^2 } }} - 1))\} \\
&= \det A \cdot \frac{1}{{\cosh \sqrt {u^2 } }} = \det A \cdot \sqrt {1 - 
(\tilde D_b \phi  + \tilde D_b \theta \gamma ^0 \lambda  - \theta \gamma 
^0 \tilde D_b \lambda )} 
\tag{29}
\end{align*}
By using the Nambu-Goto type p=2 brane action $- T\int {d^3 x} \det e$, 
and considering Eq.(29), 
it allows us to introduce a gauge field strength vector $F^r$[10] by 
variation 
of this action with respect to the field $\phi$. It is defined as 
$$
F^r  = \det A \cdot u^a  \cdot A_a^{ - 1 r}  \cdot \frac{{\sinh \sqrt 
{u^2 } }}{{\sqrt {u^2 } }}.  
\eqno{(30)}
$$
Its equation of motion results the relation $\partial _r F^r  = 0$, 
which has explicitly U(1) gauge solution $A_n$, i.e.
$$         
F_{mn}  = \partial _m A_n  - \partial _n A_m 
\eqno{(31)}
$$
and $F^r$ is related to $F_{mn}$ by
$$         
F^r  = \frac{1}{2}\varepsilon ^{mnr} F_{mn}  = \frac{1}{2}\varepsilon 
^{mnr} (\partial _m A_n  - \partial _n A_m ).
\eqno{(32)}
$$
Conversely, $F_{mn}  = \varepsilon _{mnr} F^r$. In D=3 dimension, there is 
D-2=1 degree of freedom for the 
U(1) gauge field $A_n$, which compensates the degree of freedom of field   
$\phi$ in Eq.(29). Therefore, by using Eq.(30), we find
$$
\cosh \sqrt {u^2 }  = \sqrt {1 + \sinh ^2 \sqrt {u^2 } }  = \sqrt {1 + 
\frac{{F^m A_m^{\\\ b} F^n A_n^{\\\ a} \eta _{ab} }}{{\det ^2 A}}}.   
\eqno{(33)}
$$

     Introduce the Akulov-Volkov metric field $\tilde g_{mn}$, which is 
given by   
$$
\tilde g_{mn}  = A_m ^{\\\ a} A_n ^{\\\ b} \eta _{ab}.           
\eqno{(34)}
$$
It has the explicit form
\begin{align*}
\tilde g_{mn}  = & \eta _{mn}  + i\partial _n \theta \gamma ^0 \gamma _m 
\theta  + i\partial _n \lambda \gamma ^0 \gamma _m \lambda  + i\partial _m 
\theta \gamma ^0 \gamma _n \theta  + i\partial _m \lambda \gamma ^0 \gamma 
_n \lambda \\
&+ (i\partial _m \theta \gamma ^0 \gamma ^a \theta  + i\partial _m \lambda 
\gamma ^0 \gamma ^a \lambda ) \cdot (i\partial _n \theta \gamma ^0 \gamma 
_a \theta  + i\partial _n \lambda \gamma ^0 \gamma _a \lambda ).
\tag{35} 
\end{align*}
After explicitly expanding the following determent, it can be shown
\begin{align*}
\det(\tilde g_{mn}+F_{mn})& = \det(\tilde g_{mn}+ \varepsilon _{mnr} F^r) 
\\
& = \det{\tilde g}+F^m F^n \tilde g_{mn}= {\det}^2 A \cdot {\cosh^2 \sqrt 
{u^2} }
\tag{36}
\end{align*}    
where $\tilde g = \det \tilde g_{mn}$, and the last equality is a result 
of Eq.(33). Consider the 
alternative form of Eq.(29) 
\begin{align*}                             
\det e &= \det A \cdot \frac{1}{{\cosh \sqrt {u^2 } }} = \det A \cdot 
\frac{{\cosh ^2 \sqrt {u^2 }  - \sinh ^2 \sqrt {u^2 } }}{{\cosh \sqrt {u^2 
} }} \\ 
 & = \det A \cdot \cosh \sqrt {u^2 }  - \det A \cdot \sinh \sqrt {u^2 } 
\tanh \sqrt {u^2 },  
\tag{37}
\end{align*}
by using Eqs.(19) and (30) and substituting Eq.(36) into (37), the 
resulting 
expression is
$$
\det e = \sqrt {\det (\tilde g_{mn}  + F_{mn} )}  + F^m (\partial _m \phi  
+ \partial _m \theta \gamma ^0 \lambda  - \theta \gamma ^0 \partial _m 
\lambda ) 
\eqno{(38)}
$$
   Introduce an intrinsic tensor field $ G_{mn}$, which has Weyl scale 
transformation property
$$         
G_{mn}  \to e^{2d} G_{mn}.
\eqno{(39)}
$$
Hence a spacetime integral of the first part of Eq.(38) has the 
classically equivalent Weyl invariant form: 
$$                                                                       
\int {d^3 x\{ } \tilde g^{1/4} G^{1/4} [\frac{1}{3}G^{mn} (\tilde 
g_{mn}  - \tilde g^{kl} F_{mk} F_{ln } )]^{3/4} \} 
\eqno{(40)}
$$
The equation of motion of the intrinsic tensor field $G_{mn}$, which can 
be derived from Eq.(43), is 
$$
G_{mn}  = \Omega (\tilde g_{mn}  - \tilde g^{kl} F_{mk} F_{ln } ),
\eqno{(41)}
$$   
where $\Omega$ is a constant. The spacetime integral of the second part 
of Eq.(38) has the form
\begin{align*}                                     
&\int {d^3 x} F^m (\partial _m \phi  + \partial _m \theta \gamma ^0 
\lambda  - \theta \gamma ^0 \partial _m \lambda ) \\ 
  = &\int {d^3 x} F^m (\partial _m \theta \gamma ^0 \lambda  - \theta 
\gamma ^0 \partial _m \lambda ) \\ 
= &\int {d^3 x} \frac{1}{2}\varepsilon ^{mnr} (\partial _m A_n  - \partial 
_n A_m ) \cdot (\partial _m \theta \gamma ^0 \lambda  - \theta \gamma ^0 
\partial _m \lambda ).
\tag{42}
\end{align*}
In the first equality we use the relation $\partial _r F^r  = 0$ and 
integrate by parts to drop 
the field $\phi$ term. Consider the world volume element $d^3 x$, which is 
a tensor density with weight -1. Therefore $d^3 x\sqrt {\tilde g}$ becomes 
a scalar quantity. Since $\varepsilon ^{mnr}$ is 
also a weight -1 tensor density, we can form an ordinary contravariant 
rank three tensor $\frac{{\varepsilon ^{mnr} }}{{\sqrt {\tilde g} }}$. 
Under the Weyl scale transformation $\tilde g_{mn}  \to e^{2d} \tilde 
g_{mn}$, Eq. (42) hence keeps invariant.

    Considering Eqs.(40) and (42), the dual form action, i.e. the Weyl 
scale 
invariant D2 brane Born-Infeld type action is then constructed 
\begin{align*}                            
I = & - T\int {d^3 x\{ } \tilde g^{1/4} G^{1/4} [\frac{1}{3}G^{mn} (\tilde 
g_{mn}  - \tilde g^{kl} F_{mk} F_{ln } )]^{3/4}  \\ 
 & + \frac{1}{2}\varepsilon ^{mnr} (\partial _m A_n  - \partial _n A_m ) 
\cdot (\partial _m \theta \gamma ^0 \lambda  - \theta 
\gamma ^0 \partial _m \lambda )\}             
\tag{43}
\end{align*}
where $F_{mn}$ and $\tilde g_{mn}$ are given by Eq.(31) and (35) 
respectively.

\vspace{5pt}
\begin{flushleft}
{\large IV. Localized Matter Fields on the Brane}
\end{flushleft}
\vspace{5pt}

      In addition to the massless Nambu-Goldstone fields
$\phi(x),\theta(x)$ and $\lambda(x)$ on the 
brane, there can also be matter field degrees of
freedom localized on the brane. The induced localization of the scalar and
fermionic degrees of freedom on the
submanifold were considered in [10] when the embedded defects 
spontaneously
break the target manifold. In the following
model, by using the ingredients of the Cartan one-forms, we present the 
actions of the
matter fields as well as interactions with
the Nambu-Goldstone fields. Consider there is different
dilatation scale associated with local
spacetime points on the brane, i.e.$d$ is a local function of spacetime, 
a Weyl gauge field $B_m(x)$ is introduced as the
compensating field in order to keep the whole action invariant. The action
of the Weyl gauge field $B_m(x)$  interacting with
the Nambu-Goldstone fields on the brane is also constructed.

     For the matter degrees of freedom localized on the p=2 brane, under 
the unbroken subgroup $H$, in the tangent space
the matter field $M(x)$ transforms as
  $$  M'(x') = hM(x) \\
  \eqno{(44)}
$$
in which $h$ is given by Eq.(10). The covariant derivative for the matter
field is given through the spin connection and
dilatation one-forms:
\begin{align*}
\nabla M(x) &= (d + i\omega _M^a \Gamma _a  + i\omega _D \Gamma 
^\prime)M(x)\\
&=(d + i\omega _M^a \Gamma _a  )M(x)
  \tag{45}
\end{align*}
where $\Gamma_a$ is the representation of $M_m$ corresponding to field 
$M(x)$. When 
Weyl scale
parameter $d$ becomes local function of
spacetime $d=d(x)$, Eq.(45) will not transform covariantly under $H$. A 
new 
compensating
field $B_m$  (Weyl gauge field) is introduced
[21-24], therefore in component forms the covariant derivative is written 
as:
\begin{align*}
\nabla _a M(x) &= (e_a ^{ - 1m} \partial _m  + i\omega _a^{\hspace{3pt} 
b}
\Gamma _b  + B_a d_s )M(x) \\
&=(D_a  + i\omega _a^{\hspace{3pt} b}
\Gamma _b  + B_a d_s )M(x) 
  \tag{46}
\end{align*}
where $D_a=e_a ^{ - 1m} \partial _m$, the coefficients $\omega_a^b$ 
is related to spin connection one-from by 
$\omega^b_M=\omega^{\hspace{3pt} b}_adx^a$,
$B_m$ is related to gauge field $B_a$  in the local tangent
space by $B_m=e^{\\\ a}_m B_a$, and $d_s$  is the scale dimension(weight) 
of the 
matter field $M(x)$. The
variation of the matter field under the
group $H$ is
\begin{align*}
\delta M(x) &= M'(x') - M(x) = \delta _L 'M(x) + \delta _D 'M(x) \\
&= i\varepsilon ^a  \cdot (\Gamma _a  + L_a )M(x) + id(x) \cdot DM(x) 
\tag{47}
\end{align*}
where $\delta _L 'M(x)=i\varepsilon ^a  \cdot (\Gamma _a  + L_a )M(x)$ 
represents variation under SO(1,2) transformation, the parameter 
$\varepsilon^a$
decided by Eq.(10) is a function of $\alpha$ and $b$. And $L$
is the angular momentum representation of $M_m$.  The variation of the 
scale
transformation is
\begin{align*}
 \delta _D 'M(x) &= M'(x') - M(x)|_D  =  id(x) \cdot DM(x) \\
 &= d(x \cdot \partial  + d_s )M(x) 
  \tag{48}
\end{align*}
Hence, the general filed representation of the scale operator is given by
$$  D = -i(x \cdot \partial  + d_s ). \\
\eqno{(49)}
$$
The intrinsic Weyl scale variation of the matter field then can be written 
as
  $$  \delta _D M(x) = M'(x) - M(x)|_D  = d(x) \cdot d_s M(x). \\
  \eqno{(50)}
$$
From Eqs.(18) and (25) one can find the intrinsic infinitesimal scale 
variation 
of
dreibein
  $$  \delta _D e_m^{\\\ a}  =  de_m^{\\\ a}  \\
  \eqno{(51)}
$$
Therefore the scale dimension for dreibein is 1. Besides, in 2+1 
dimensions,
the scalar
field has weight $d_s (\phi)=-1/2$, and the spinor field is 
$d_s(\psi)=-1$. 
Correspondingly, we have
\begin{align*}
  \delta _D \phi  &=  - \frac{1}{2}d(x)\phi,  \\
 \delta _D \psi  &= \delta _D \bar \psi  =  - d(x)\psi,  
\tag{52}
\end{align*}
with scale transformation of the coordinates in the tangent space
  $$  \delta _D x^a  = d(x)x^a.  \\
  \eqno{(53)}
$$
Hence, under $D$ operation the derivative of the matter field transforms 
as
\begin{align*}
  D_a M(x) &\to D'_a M'(x') = D'_a M'(D x) \\
 &  = e^{ - d(x)} e_a^{ - 1m} \partial _m e^{d(x)d_s } M(x) \\
&  = e^{ - d(x)} e^{d(x)d_s } e_a^{ - 1m} (d_s d(x)_{,m} M(x) + \partial
_m M(x)) 
\tag{54}
\end{align*}
The Weyl gauge covariant derivative then transforms as
\begin{align*}
   \nabla _a M(x) &\to \nabla '_a M'(x') = (D'_a  + i\omega 
_a^{\hspace{3pt} b} \Gamma
_b  + B_a 'd_s )M'(Dx) \\
&  = e^{-d(x)} e^{d(x)d_s } (D_a  + d_s d(x)_{,a}  + i\omega 
_a^{\hspace{3pt} b} \Gamma 
_b+ B_a d_s  - d_s d(x)_{,a} )M(x) \\
&  = e^{-d(x)} e^{d(x)d_s } \nabla _a M(x) 
  \tag{55}
\end{align*}
in which we used $\delta_D \omega _M^b=0$, therefore $\delta_D \omega 
_a^{\hspace{3pt} b}=-d(x)\omega_a^{\hspace{3pt} b}$ and the Weyl gauge 
field $B_m$  has the 
infinitesimal scale transformation property
 $$  \delta _D B_m (x) =  - d(x)_{,m}.  \\
  \eqno{(56)}
$$

    Considering the scalar field localized on the p=2 brane, since 
$\Gamma_b(\phi)=0$  and
$d_s(\phi)=-1/2$, the Weyl gauge covariant derivative is then constructed
  $$  \nabla _a \phi (x) = (e_a^{ - 1m} \partial _m  - \frac{1}{2}e_a^{ -
1m} B_m )\phi (x). \\
  \eqno{(57)}
$$
The Lagrangian density of the scalar field is given by
\begin{align*}
  \ell _\phi   &= \eta ^{ab} \nabla _a \phi (x)\nabla _b \phi (x) +
f\phi ^6  \\
&= g^{mn} (\partial _m  - \frac{1}{2}B_m )\phi (x)(\partial _n  -
\frac{1}{2}B_n )\phi (x) + f\phi ^6  
\tag{58}
\end{align*}
in which $f$ is the dimensionless coupling constant. The effective action 
of the scalar matter field on the brane up to
the leading term in brane tension expansion is obtained
$$  I_\phi   = \int {d^3 x} \det e\ell _\phi.   \\
\eqno{(59)}
$$
For the fermion spinor field, the spinor representation of the operators
 $M_a$ are $\Gamma_a(\psi)=-\frac{1}{2}\gamma_a$ . The covariant 
derivative of the spinor field is
  $$  \nabla _a \psi _i (x) = D_a \psi _i  - i\frac{1}{2}\omega 
_a^{\hspace{3pt} b}
\gamma _{bij} \psi _j (x) - B_a \psi _i (x) \\
  \eqno{(60)}
$$
Since $\psi(x),\bar \psi(x)$  interacts with the field $B_a$ in the 
same manner, the spinor field has
no minimal form of coupling to the Weyl
gauge field. The Lagrangian of the spinor matter field with Yukawa 
coupling
to the scalar fields is
  $$  \ell _\psi   = \frac{1}{2}i[\bar \psi \gamma ^a \nabla _a \psi  -
\nabla _a \bar \psi \gamma ^a \psi ] + g\bar \psi \psi \phi ^2  \\
  \eqno{(61)}
$$
The effective action of the spinor field on the brane has the form
  $$  I_\psi   = \int {d^3 x} \det e\ell _\psi   \\
  \eqno{(62)}
$$
The field strength which describes the Weyl gauge field $B_m$ has the 
normal 
form
  $$  F_{mn}  = \partial _m B_n  - \partial _n B_m  \\
  \eqno{(63)}
$$
Introducing new dynamics variables $F_{ab}=e^{-1m}_ae^{-1n}_bF_{mn}$, 
their infinitesimal Weyl transformation properties are
  $$  \delta _D F_{ab}  =  - 2d(x)F_{ab},  \\
  \eqno{(64)}
$$
on dimension and Weyl scale invariant ground, the effective action of the 
Weyl gauge field can be constructed
  $$  I_B  = \int {d^3 x} \det e\ell _B  = \int {d^3 } x\det e\ell_B 
(e^{-1m}_a, F_{mn})  \\
  \eqno{(65)}
$$
where the lagrangian has Weyl dimension -3 and is a function of 
$e_a^{-1m}$ and Weyl gauge field strength $F_{mn}$. Considering 
Eqs.(59),(62) and (65), the full effective action for the matter fields
localized on the brane is then given by
\begin{align*}
   I_{matter}  &= I_\phi   + I_\psi   + I_B  = \int {d^3 x} \det e\ell
_{matter}  \\
  &= \int {d^3 x} \det e\ell _\phi   + \int {d^3 x} \det e\ell _\psi   +
\int {d^3 x} \det e\ell _B  
  \tag{66}
\end{align*}

    In summary, in this letter we have constructed Weyl scale invariant
version of the p=2 brane action, which is a
result of spontaneous breaking of the target N=1, D=4 superspace with
$G$ symmetry to the $W(1,2)$ symmetry on the embedded the
2+1 world volume.  Its low energy fluctuations in directions associated
with the broken symmetry generators are
described by the dynamics of the Nambu-Goldstone fields. There, unlike the
BPS state of the D brane which carries
conserved charges or the partially broken supersymmetry on the brane whose
central charge saturates the lower bound of
the state [25], it is the case of non-BPS state. By this approach of
nonlinear realization, one can also find its
application to branes of M theory with a large automorphism group of
superalgebra [26]. In addition, as described
above, the brane, as a defect in spacetime that breaks certain symmetries,
may cause the localization of matter fields
as well as the gauge fields on it, which is a fact of physical necessity 
and required to be present in the effective
world volume field theories. Additional discussions can also be found
in [27, 28] and some brane world scenarios as
well [29].

The author thanks T.K.Kuo for support and discussions. The author also
thanks M.Burkardt and Y.X.Gui for their kind support. Finally, the author 
would like to acknowledge helpful discussions for the revising version of 
the manuscript in TASI 2006. 

\pagebreak

\begin{flushleft}
{\large Appendix A: Dimensional Reduction of N=1, D=4 SUSY to N=2, D=3 
SUSY}
\end{flushleft}
\vspace{5pt}

   In N=2, D=3 Supersymmetry theory, there are two two-component Majorana
spinorial generators $Q_\alpha ^1,Q_\beta^2$  satisfying
\begin{align*}
   \{ Q_\alpha ^1 ,\bar Q_\beta ^1 \}  &= (\gamma  \cdot P)_{\alpha
\beta },  \\
 \{ Q_\alpha ^2 ,\bar Q_\beta ^2 \}  &= (\gamma  \cdot P)_{\alpha \beta },
\\
 \{ Q^1 ,\bar Q^2 \} &= iZ_0,
  \tag{{\it{A}}.1}
\end{align*}
where $\alpha,\beta=1,2$  are component indices,  and $ (\gamma ^0 
,\gamma ^1 ,\gamma ^2 ) = (\sigma ^2 ,i\sigma ^1 ,i\sigma ^3 )$,$ \bar Q 
^{1(2)}= Q^{1(2) \dagger}  C = Q^{1(2) \dagger}  \gamma ^0 $  , with 
Majorana condition $Q^{1(2)} =  - Q^{1(2)*}  $ , and $Z_0$ is the central 
charge. If we 
introduce a Dirac spinor $ Q' = \frac{1}{{\sqrt 2 }}(Q^1  + iQ^2 )$, 
then we have
  $$  \{ Q',\bar Q'\}  = \gamma  \cdot P + Z_0.  \\
  \eqno{(A.2)}
$$

    In N=1, D=4 Supersymmetry theory, from Eq.(1), we have
  $$  \{ Q_\alpha  ,\bar Q_{\dot \alpha } \}  = 2\sigma _{\alpha \dot
\alpha }^\mu  P_\mu.   \\
  \eqno{(A.3)}
$$
Imposing rotation operations $e^{ - iM^{23} \frac{\pi }{2}}$ and 
$e^{ iM^{13} \frac{\pi }{2}}$ consecutively on Eq.(A.3), the four 
momentum vector has the corresponding transformation
\begin{align*}
P_0 &\to P_0\\
P_1 &\to P_2\\
P_2 &\to -P_3\\
P_3 &\to -P_1
  \tag{{\it{A}}.4}
\end{align*}
and the spinor becomes
  $$Q \to W\\
  \eqno{(A.5)}
$$
with
$$W = \left( \begin{array}{ccc}
{Q_1 e^{ - i\frac{\pi }{4}} \cos \frac{\pi }{4} + Q_2 e^{i\frac{\pi
}{4}} \sin \frac{\pi }{4}}  \\
{Q_2 e^{i\frac{\pi }{4}} \cos \frac{\pi }{4} - Q_1 e^{ - i\frac{\pi
}{4}} \sin \frac{\pi }{4}}  \\
\end{array} \right)
\eqno{(A.6)}
$$
Equation (A.3) then has the form
$$  \{ W,W^{\dagger}   \}  = 2\sigma ^0 P_0  + 2\sigma ^1 P_2  + 2\sigma 
^2 ( -
P_3 ) + 2\sigma ^3 ( - P_1 ) \\
\eqno{(A.7)}
$$
Right multiplication of $\sigma^2$ from both sides, it has the form  
$$  \{ W,W^{\dagger}  \sigma ^2 \}  = 2\sigma ^2 P_0  + 2i\sigma ^3 P_2  + 
2( -
P_3 ) + 2i\sigma ^1 P_1  \\$$
$$  = 2\gamma ^0 P_0  + 2\gamma ^1 P_1  + 2\gamma ^2 P_2  + 2( - P_3 ). \\
\eqno{(A.8)}
$$
Thus
$$  \{ W,\bar W\}  = 2\gamma  \cdot P + 2( - P_3 ), \\
\eqno{(A.9)}
$$
where $\bar W=W^{\dagger} \gamma^0$. Compare (A.9) with (A.2), we may 
identify $W$ 
with $\sqrt{2}Q'$ and $Z_0$  
with $-P_3$. Using
redefined operators
$$  Q^1  = \frac{1}{{\sqrt 2 }}\left( \begin{array}{ccc}
   {is_1 }  \\
   {is_2 }  \\
\end{array} \right)
, Q^2  = \frac{1}{{\sqrt 2 }}\left( \begin{array}{ccc}
   { - iq_1 }  \\
   { - iq_2 }  \\
\end{array} \right),
\eqno{(A.10)}
$$
from Eqs.(A.1), (A.2), (A.6) and (A.10), we have
$$  \left( {\begin{array}{*{20}c}
   {q_1 }  \\
   {q_2 }  \\
\end{array}} \right) = \left( {\begin{array}{*{20}c}
   {\frac{1}{2}Q_1 e^{i\frac{\pi }{4}}  - \frac{1}{2}Q_2 e^{i\frac{\pi
}{4}}  + \frac{1}{2}\bar Q_1 e^{ - i\frac{\pi }{4}}  - \frac{1}{2}\bar Q_2
e^{ - i\frac{\pi }{4}} }  \\
   {\frac{1}{2}Q_1 e^{ - i\frac{\pi }{4}}  + \frac{1}{2}Q_2 e^{ -
i\frac{\pi }{4}}  + \frac{1}{2}\bar Q_1 e^{i\frac{\pi }{4}}  +
\frac{1}{2}\bar Q_2 e^{i\frac{\pi }{4}} }  \\
\end{array}} \right), \\$$
$$
 \left( {\begin{array}{*{20}c}
   {s_1 }  \\
   {s_2 }  \\
\end{array}} \right) = \left( {\begin{array}{*{20}c}
  {\frac{1}{2}Q_1 e^{ - i\frac{\pi }{4}}  - \frac{1}{2}Q_2 e^{ -
i\frac{\pi }{4}}  - \frac{1}{2}\bar Q_1 e^{i\frac{\pi }{4}}  +
\frac{1}{2}\bar Q_2 e^{i\frac{\pi }{4}} }  \\
   {\frac{1}{{2i}}Q_1 e^{ - i\frac{\pi }{4}}  + \frac{1}{{2i}}Q_2 e^{ -
i\frac{\pi }{4}}  - \frac{1}{{2i}}\bar Q_1 e^{i\frac{\pi }{4}}  -
\frac{1}{{2i}}\bar Q_2 e^{i\frac{\pi }{4}} }  \\
\end{array}} \right). \\
\eqno{(A.11)}
$$
Hence, the extended centrally charged N=2, D=3 supersymmetry algebra is
$$
\{ q_i ,q_j \}  = 2(\gamma ^m C)_{ij} P_m,
 \{ s_i ,s_j \}  = 2(\gamma ^m C)_{ij} P_m,
 \{ q_i ,s_j \}  =  - 2iC_{ij} P_3, $$ 
$$[K^m ,q_j]   = \frac{1}{2}\gamma ^m_{ij} s_j,
[K^m ,s_i ]  = -\frac{1}{2}\gamma ^m_{ij} q_j,$$
$$[M^{mn},q_i]  = -\frac{1}{2}\gamma ^{mn}_{ij} q_j,
[M^{mn},s_i]  = -\frac{1}{2}\gamma ^{mn}_{ij} s_j.
\eqno{(A.12)}
$$

\vspace{5pt}
\begin{flushleft}
{\large Appendix B: Useful Formulas}
\end{flushleft}
\vspace{5pt}

   In derivation of Eqs.(9) and (10), we consider the Baker-Hausdorff 
formula:
$$  \exp (a)\exp (b) = \exp (a + b + \frac{1}{2}[a,b] +
\frac{1}{{12}}[a,[a,b]] + \frac{1}{{12}}[b,[b,a]] + ...). \\
\eqno{(B.1)}
$$
For infinitesimal operator $a$, up to the first order of $a$, we have
$$  \exp (a)\exp (b) = \exp (a + b + \frac{1}{2}[a,b] +
\frac{1}{{12}}[b,[b,a]] + ... + O(a^2 )) \\$$
$$  = \exp (a - ad_{b/2} (a) + ad_{b/2}  \cdot \coth (ad_{b/2} )(a) + 
O(a^2)); \\
\eqno{(B.2)}
$$
and likewise
$$  \exp (b)\exp (a) = \exp (a + b + \frac{1}{2}[b,a] +
\frac{1}{{12}}[b,[b,a]] + ...) \\$$
$$  = \exp (b + ad_{b/2} (a) + ad_{b/2}  \cdot \coth (ad_{b/2} )(a) + 
O(a^2)), \\
\eqno{(B.3)}
$$
in which $ad_{b/2}(a)$ is the adjoint operation with 
$ad_{b/2}(a)=[\frac{b}{2},a]$. In derivation of 
Eq.(16),consider the differentiation formula for
exponent:
$$  \exp ( - b)d\exp (b) = \sum\limits_{k = 0}^\infty  {\frac{{( - 1)^k
}}{{(k + 1)!}}} (ad_b )^k db \\
\eqno{(B.4)}
$$                                                                      

\pagebreak

\begin{center}
{\bf REFERENCES}
\end{center}
\begin{description}

\item[[1]] S.Coleman, J.Wess and R.Zumino, Phys. Rev. 177, 2239(1969);
Phys. Rev. 177,
     2247(1969)

\item[[2]] D. V. Volkov and V. P. Akulov, JETP Lett. 16, 438(1972);
Phys. Lett. B46,
     109(1973)

\item[[3]] Lu-Xin Liu, Mod. Phys. Lett. A20, 
2545(2005)[arXiv:hep-ph/0408210] 

\item[[4]] T. E. Clark and S. T. Love and G. Wu, Phys.Rev.D57, 5912(1998)
\item[[5]] P. West, JHEP 0002, 24(2000)

\item[[6]] J. Hughes and J. Polchinski, Nucl. Phys. B278, 147(1986);
C. P. Burgess, E. Filotas,
     M. Klein and F. Quevedo, JHEP 0310, 41(2003); J. Bagger, A. Galperin,
Phys. Lett. B
     336, 25(1994); Phys. Lett. B412, 296(1997); Phys. Rev. D55, 
1091(1997)

\item[[7]] S. Bellucci, E. Ivanov and S. Krivonos, Phys. Lett. B460,
348(1999); M. Rocek and
      A.A.Tseytlin, Phys. Rev. D59, 106001(1999); S. Bellucci, E. Ivanov
and S. Krivonos,
      Fortsch.Phys.48, 19(2000); F. Gonzalez-Rey, L. Y. Park, and M. 
Rocek,
Nucl. Phys.
      B544, 243(1999)

\item[[8]] E.Ivanov and S.Krivonos, Phys. Lett. B453, 237(1999)

\item[[9]] I. N. McArthur, Hep-th/9908045; S. Bellucci, E. Ivanov,
S. Krivonos, Phys.Lett.B
     482, 233(2000); S. Bellucci, E. Ivanov and S. Krivonos, Nucl. Phys.
Proc. Suppl. 102,
     26(2001); E. Ivanov, Theor. Math. Phys. 129, 1543(2001); E. Ivanov, 
AIP Conference Proceedings 589, 61(2001)

\item[[10]] T. E. Clark, M. Nitta and T. ter Veldhuis, Hep-th/0208184

\item[[11]] M. S. Alver and J. Barcelos-Neto, Europhys. Lett. 7, 
395(1988);
Erratum, Europhys.
      Lett. 8, 90(1989); C. Alvear, R. Amorim and J. Barcelos-Neto, Phys.
Lett. B273, 415(1991)
\item[[12]] J. Antonio Garcia, R. Linares and J. David Vergara, Phys. 
Lett.
B503, 154(2001)

\item[[13]] See, for example, F. Delduc, E. Ivanov and S. Krivonos, Phys.
Lett. B529, 233
       (2002)

\item[[14]] M. Blagojevic, Gravitation and Gauge Symmetries (Institute of
Phys. Publishing, London,
       2002)

\item[[15]] D. G. C. Mckeon and T. N. Sherry, hep-th/0108074

\item[[16]] D. G. C. McKeon, Nucl. Phys. B591, 591(2000)

\item[[17]] Taichiro Kugo and Paul Townsend, Nucl. Phys, B221, 357(1983)

\item[[18]] S. Weinberg, The Quantum Theory of Fields (Cambridge Univ.
Press, 1996), Vol. 2

\item[[19]] R. Casalbuoni, About the counting of the Goldstone bosons,
Seminar in Florence,
       2001

\item[[20]] E. A. Ivanov and V. I. Ogievetsky, Teor. Mat. Fiz. 25, 
164(1975)

\item[[21]] K. Hayashi, M. Kasuya and T.Shiraguji, Progress of Theo.
Phys. 57, 431(1977)

\item[[22]] A. Bregman, Pro. Theo. Phys. 49, 667(1973)

\item[[23]] R. UtiYama, Pro. Theo. Phys. 50, 2080(1973)

\item[[24]] K. Hayashi and T.Kugo, Prog. Theo. Phys. 61, 334(1979)

\item[[25]] A. Kovner, M. Shifman and A. Smilga, Phys. Rev. D56, 
7978(1997)

\item[[26]] O. Barwald and P. West, Phys. Lett. B476, 157(2000)

\item[[27]] T. E. Clark, M. Nitta and T.ter Veldhuis, Hep-th/0209142

\item[[28]] G. Dvali and M. Shifman, Phys. Lett. B396, 64(1997);
Erratum-ibid.B407,
       452(1997); E. K. Akhmedov, Phys. Lett. B521, 79(2001)

\item[[29]] K. Akama, Hep-th/0001113

\end{description}

\end{document}